\documentclass[twocolumn, prb, aps, superscriptaddress, showkeys]{revtex4-1}

\usepackage{graphicx}
\usepackage{amsmath}
\usepackage{bm}
\usepackage{xcolor}
\usepackage{color}
\usepackage{siunitx}
\usepackage{physics}
\usepackage{natmove}
\usepackage[colorlinks=true,
            linkcolor = blue,
            urlcolor  = blue,
            citecolor = blue,
            anchorcolor = blue]{hyperref}
\usepackage{tikz}
\newcommand{\beq}{\begin{equation}}
\newcommand{\eeq}{\end{equation}}
\newcommand{\bea}{\begin{eqnarray}}
\newcommand{\eea}{\end{eqnarray}}

\begin{document}

%\title{Functional Form of the Superconducting Critical Temperature from Machine Learning}
\title{Machine learning of superconducting critical temperature from Eliashberg theory}
\author{S.~R. Xie}
\affiliation{Department of Materials Science and Engineering,  University of Florida, Gainesville FL 32611, USA}
\affiliation{Quantum Theory Project, University of Florida, Gainesville FL 32611, USA}
\author{Y. Quan}
\affiliation{Department of Materials Science and Engineering,  University of Florida, Gainesville FL 32611, USA}
\affiliation{Quantum Theory Project, University of Florida, Gainesville FL 32611, USA}
\affiliation{Department of Physics, University of Florida, Gainesville, Florida 32611, USA}
\author{A.\ C.\ Hire}
\affiliation{Department of Materials Science and Engineering,  University of Florida, Gainesville FL 32611, USA}
\affiliation{Quantum Theory Project, University of Florida, Gainesville FL 32611, USA}

\author{B. Deng}
\affiliation{Department of Physics, University of Florida, Gainesville, Florida 32611, USA}
\author{J.\ M.\ DeStefano}
\affiliation{Department of Physics, University of Florida, Gainesville, Florida 32611, USA}
\author{I. Salinas}
\affiliation{Department of Physics, University of Florida, Gainesville, Florida 32611, USA}
\author{U. S. Shah}
\affiliation{Department of Physics, University of Florida, Gainesville, Florida 32611, USA}
\author{L.\ Fanfarillo}
\affiliation{Department of Physics, University of Florida, Gainesville, Florida 32611, USA}
\affiliation{Scuola Internazionale Superiore di Studi Avanzati (SISSA), Via Bonomea 265, 34136 Trieste, Italy}
\author{J.\ Lim}
\affiliation{Department of Physics, University of Florida, Gainesville, Florida 32611, USA}
\author{J.\ Kim}
\affiliation{Department of Physics, University of Florida, Gainesville, Florida 32611, USA}
\author{G.~R. Stewart}
\affiliation{Department of Physics, University of Florida, Gainesville, Florida 32611, USA}
\author{J.~J. Hamlin}
\affiliation{Department of Physics, University of Florida, Gainesville, Florida 32611, USA}
\author{P.~J. Hirschfeld}
\affiliation{Department of Physics, University of Florida, Gainesville, Florida 32611, USA}
\author{R.~G. Hennig}
\email{rhennig@ufl.edu}
\affiliation{Department of Materials Science and Engineering,  University of Florida, Gainesville FL 32611, USA}
\affiliation{Quantum Theory Project, University of Florida, Gainesville FL 32611, USA}
\date{\today}
\begin{abstract}
% We build on earlier efforts by McMillan and Allen and Dynes to model $T_c$ from various measures of the phonon spectrum and the electron-phonon interaction by using machine learning to identify an analytical prefactor to the ubiquitous McMillan equation, correcting the underestimation of $T_c$ at large $\lambda$. To address the limitations of the Allen-Dynes equation arising from their selection of spectral function $\alpha^2F(\omega)$ shapes, we generate new data using Eliashberg Theory and more $\alpha^2F$ examples, ranging from multimodal Einstein-like models to calculated spectra of polyhydrides. Using symbolic regression and the sure independence screening and sparsifying operator (SISSO) framework, we identify a new equation and benchmark $T_c$ predictions alongside random forest and artificial neural network models trained on the same data. The expression identified through our data-driven approach corrects the systematic underestimation of $T_c$ while reproducing the physical constraints originally outlined by Allen and Dynes.
The Eliashberg theory of superconductivity accounts for the fundamental physics of conventional electron-phonon superconductors, including the retardation of the interaction and the effect of the Coulomb pseudopotential, to predict the critical temperature $T_c$ and other properties. McMillan, Allen, and Dynes derived approximate closed-form expressions for the critical temperature predicted by this theory, which depends essentially on the electron-phonon spectral function $\alpha^2F(\omega)$, using $\alpha^2F$ for low-$T_c$ superconductors. Here we show that modern machine learning techniques can substantially improve these formulae, accounting for more general shapes of the $\alpha^2F$ function.  Using symbolic regression and the sure independence screening and sparsifying operator (SISSO) framework, together with a database of artificially generated $\alpha^2F$ functions, ranging from multimodal Einstein-like models to calculated spectra of polyhydrides, as well as numerical solutions of the Eliashberg equations, we derive a  formula for $T_c$ that performs as well as Allen-Dynes for low-$T_c$ superconductors and substantially better for higher-$T_c$ ones.  The expression identified through our data-driven approach corrects the systematic underestimation of $T_c$ while reproducing the physical constraints originally outlined by Allen and Dynes. This equation should replace the Allen-Dynes formula for the prediction of higher-temperature superconductors and the estimation of $\lambda$ from experimental data.
%We discuss how such expressions can guide the search for higher-temperature superconductors at ambient pressure.
\end{abstract} 
\maketitle

\section{Introduction}
Although the theory of electron-phonon superconductivity due to Bardeen-Cooper-Schrieffer, Gor'kov, Eliashberg, Migdal, and others is well-established, it has not historically aided in the discovery of new superconductors.  The materials space to search for new superconductors is vast, and it is, therefore, desirable to find a practical way to use theory as a guide. Recent computational developments may allow a new approach to superconducting materials discovery based on ab-initio and materials-genome type methods~\cite{genome_SC,Pickett2018,Duan_review2019}.

One approach to this problem, pioneered by MacMillan~\cite{Mcmillan1968} and Allen and Dynes\cite{Allen-Dynes1975}, is to search for a formula for $T_c$ based on materials-specific parameters derived from the Eliashberg equations of superconductivity.  These parameters, mostly moments of the electron-phonon spectral function $\alpha^2F(\omega)$, can be determined by experiment or, more recently, calculated within \emph{ab initio} approaches. In principle, this allows one also to deduce how to optimize $T_c$ if one can optimize one or more of these parameters. 

The Allen-Dynes equation has played a crucial role in debates on how to achieve high-temperature superconductivity  by both theorists, who use it to predict $T_c$, and by experimentalists, who extract $\lambda$ from measured $T_c$ and $\omega_D$.
% and indeed is often used to extract quoted values of the dimensionless interaction parameter $\lambda$ in the literature for materials where tunneling data is not available.
% {\blue [RGH: The Allen-Dynes equation uses $\lambda$ and emphasizes its importance in achieving high $T_c$ but does not extract this number. We should reword this sentence.]}
Nevertheless, it is important to recall that the Allen-Dynes equation has been derived from Eliashberg theory within an approximation where the momentum dependence of the Eliashberg function is neglected. It is based on 217 Eliashberg solutions of three types of $\alpha^2F(\omega)$ {\it shapes} (those obtained from tunneling data on Hg and Pb, and those obtained for a  single Einstein mode).

There have been several important advances in providing more detailed solutions to the Eliashberg equations since the work of Allen and Dynes.  Combescot solved the Eliashberg equations on the weak coupling side and obtained an expression for T$_c$ that depends on $\expval{\omega_{\log}}$ and a shape-dependent integral~\cite{PhysRevB.42.7810}.
Recently, Marsiglio {\it et al.} solved the Eliashberg equations at small $\lambda$ and found a correction factor of $\frac{1}{\sqrt{e}}$ to the BCS $T_c$~\cite{PhysRevB.98.024523, PhysRevB.101.064506}. And of course the full equations can be solved numerically, including the momentum dependence of $\alpha^2F$ if desired~\cite{EPW-2, Margine2013}.

In this paper, we solve the Eliashberg equations using different types of electron-phonon spectral functions, including multimodal Einstein-like spectra and a set of $\alpha^2F$ obtained from  first-principles calculations. We find that, while the Allen-Dynes formula accurately predicts the Eliashberg $T_c$ for $\lambda$ values near 1.6 (the coupling constant for Hg and Pb), it nevertheless deviates from the Eliashberg $T_c$ when $\lambda$ is significantly larger or smaller than 1.6 and when the shape of $\alpha^2F(\omega)$ differs from the simple unimodal Einstein model.  This deficiency highlights the need to improve on Allen-Dynes to investigate the high-pressure, high-temperature hydrides of great current interest~\cite{Flores-Livas2020}.

In a previous paper, we used an analytical machine learning approach to try to improve on the Allen-Dynes formula, testing and training on tiny databases from the Allen-Dynes table of 29 superconducting materials~\cite{Xie2019}. This proof of principle work showed that the SISSO framework, properly constrained by physical law, could substantially improve the performance of the Allen-Dynes equation with a smaller number of parameters. 
Clearly, it is necessary to apply this approach to a more extensive and diverse database.

\begin{figure}[t]
  \includegraphics[width=\columnwidth]{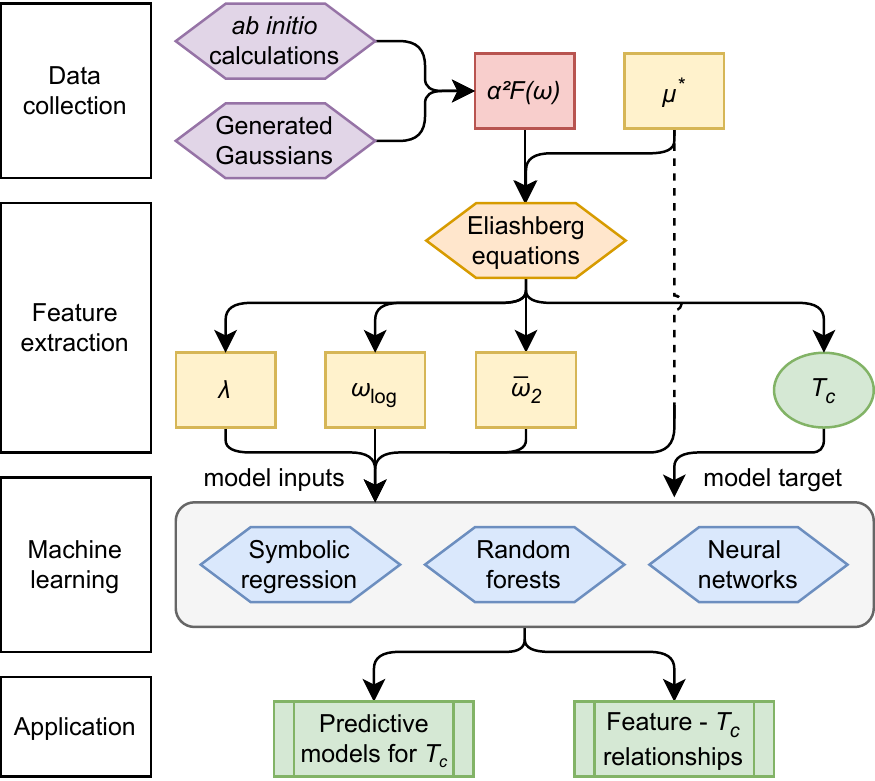}
  \caption{Workflow for identifying new machine learning models for $T_c$ from $\alpha^2F(\omega)$ spectra and derived quantities. The workflow is organized into the four computational modules listed on the left.}
  \label{fig:workflow}
\end{figure}

Here, we proceed more systematically and show how we can ``teach the machine Eliashberg theory" by generating large databases of $\alpha^2F$ functions from both real materials and single- and multimodal artificial ones and learning the results of $T_c$ from solutions to the Eliashberg equations.  We additionally include in our study $\alpha^2F$ functions for superhydrides, extending training and testing to the higher $\lambda$ range.  We show that the Allen-Dynes equation fails in this region particularly badly, since it was designed to fit materials with the ratio of the Allen-Dynes parameters  $\bar{\omega}_2/\omega_{\log}\simeq 1$, which is strongly violated in some of the higher-$T_c$ materials. Here  $\lambda$ is the integral $2\int_0^\infty \alpha^2F(\omega)/\omega\, d\omega$,  the frequencies $\bar \omega_n$ are the $n^\mathrm{th}$ root of the $n^\mathrm{th}$ moment of the normalized distribution
 $g(\omega) = 2/(\lambda \omega) \alpha^2F(\omega)$, and  $\omega_{\log}\equiv \exp \langle \ln \omega\rangle$.

We begin by introducing the McMillan and Allen-Dynes equations, against which we will compare our results.   McMillan~\cite{Mcmillan1968}, in an attempt to improve on the BCS weak-coupling $T_c$, incorporated elements of Eliashberg theory~\cite{EliashbergInteractionBetweenElAndLatticeVibrInASC1960} into a phenomenological expression, relating $T_c$ to  physical parameters  that could in principle be extracted from tunneling data~\cite{McmillanRowell1965},
\begin{equation}
  T_c  \simeq \frac{\omega_D}{1.45}\exp\left(-\frac{1.04(1+\lambda)}{\lambda-\mu^\ast(1+0.62\lambda)} \right),\label{eq:McMillan}
\end{equation}
where $\mu^\ast$ is the Coulomb pseudopotential and $\omega_D$ is the Debye frequency. Note that the McMillan formula  predicts a saturation of $T_c$ in the strong-coupling limit, $\lambda\rightarrow \infty$, for fixed $\omega_D$.

Allen and Dynes~\cite{Allen-Dynes1975} showed that the true Eliashberg $T_c$ did not obey such a bound in this limit but rather grew as $\sqrt{\lambda}$.  They proposed an alternate approximate fit to Eliashberg theory based on data on a few low-$T_c$ superconductors known in 1975, 
\begin{equation}
T_c={\frac{f_1f_2 \omega_{\log}}{1.20}} \exp\left(-\frac{1.04(1+\lambda)}{\lambda-\mu^\ast(1+0.62\lambda)}\right), \label{eq:AD}
\end{equation}

\begin{equation}
f_1 = \left(1 + \left(\frac{\lambda}{2.46(1 + 3.8 \mu^{\ast})}\right)^{3/2}\right)^{1/3},
\label{eq:f1}
\end{equation}

\begin{equation}
f_2 = \left(1 + \frac{\lambda^2(\frac{\bar{\omega}_{2}}{\omega_{\log}} - 1)}{\lambda^2 + 1.82(1 + 6.3\mu^{\ast})(\frac{\bar{\omega}_{2}}{\omega_{\log}})^2}\right),
\label{eq:f2}
\end{equation}
where $f_1$ and $f_2$ are  factors depending on $\lambda,\mu^\ast,\omega_{\log}$, and $\bar \omega_2$.
% The additional tunneling-derived parameters $\omega_\mathrm{ph}$, defined as the high-frequency cutoff in $\alpha^2F(\omega)$, and $\eta$, defined as McMillan-Hopfield parameter, also appear in their discussion.
% They showed that the expression \eqref{eq:AD} fit the $T_c$ of  a variety of superconductors known at the time, using data derived from tunneling, and that it implied the absence of any maximum $T_c$, except that caused by the competition between $\lambda$ and $\omega_{\log}\equiv \exp \langle \ln \omega\rangle$, where the average is taken over $g(\omega)$. 
% Unlike the McMillan expression,  the Allen-Dynes  equation obeys an asymptotic result of Eliashberg theory, that $T_c\sim \sqrt{\lambda}$ as $\lambda\rightarrow \infty$. 

\section{Method and data}

Fig.~\ref{fig:workflow} outlines our methods and computational workflow. We begin by collecting $\alpha^2F(\omega)$ spectral functions from \emph{ab initio} calculations and augmenting the dataset with artificial spectral functions based on generated Gaussian functions. The Coulomb pseudopotential $\mu^\ast$ is sampled as a free parameter and used, alongside the spectral functions, as an input to the Eliashberg equations. Eliashberg theory yields the superconducting gap function $\Delta$, from which we extract $T_c^E$. At the same time, we extract the quantities $\lambda$, $\omega_{\log}$, and $\bar\omega_2$ from $\alpha^2F$. Next, we use machine learning techniques to learn the relationship between the four model inputs, or features, and the critical temperature from Eliashberg theory $T_c^E$. Finally, we compare the predictive models for $T_c$ and discuss the feature-$T_c$ relationships.

\subsection{Computational details}

We compile a set of 2874 electron-phonon spectral functions $\alpha^2F(\omega)$, summarized in Tab.~\ref{tab:data_summary}. Of these, 13 are  conventional phonon mediated superconductors, where we calculate $\alpha^2F$ using the electron-phonon Wannier package (EPW)~\cite{EPW-1, EPW-2} of the Quantum Espresso (QE) code~\cite{doi:10.1063/5.0005082, Giannozzi_2009}.
An additional 42 (29 classic and 13 hydride superconductors) are obtained from the computational superconductivity literature.  We augment the dataset by generating 2819 artificial multimodal $\alpha^2F(\omega)$ functions and calculating the corresponding T$_c$s with the EPW code.   The superconducting transition temperatures are estimated  by using both the Allen-Dynes equation and by solving the isotropic Eliashberg equations. The raw data is available upon request.

\begin{figure}[t]
  \includegraphics[width=\columnwidth]{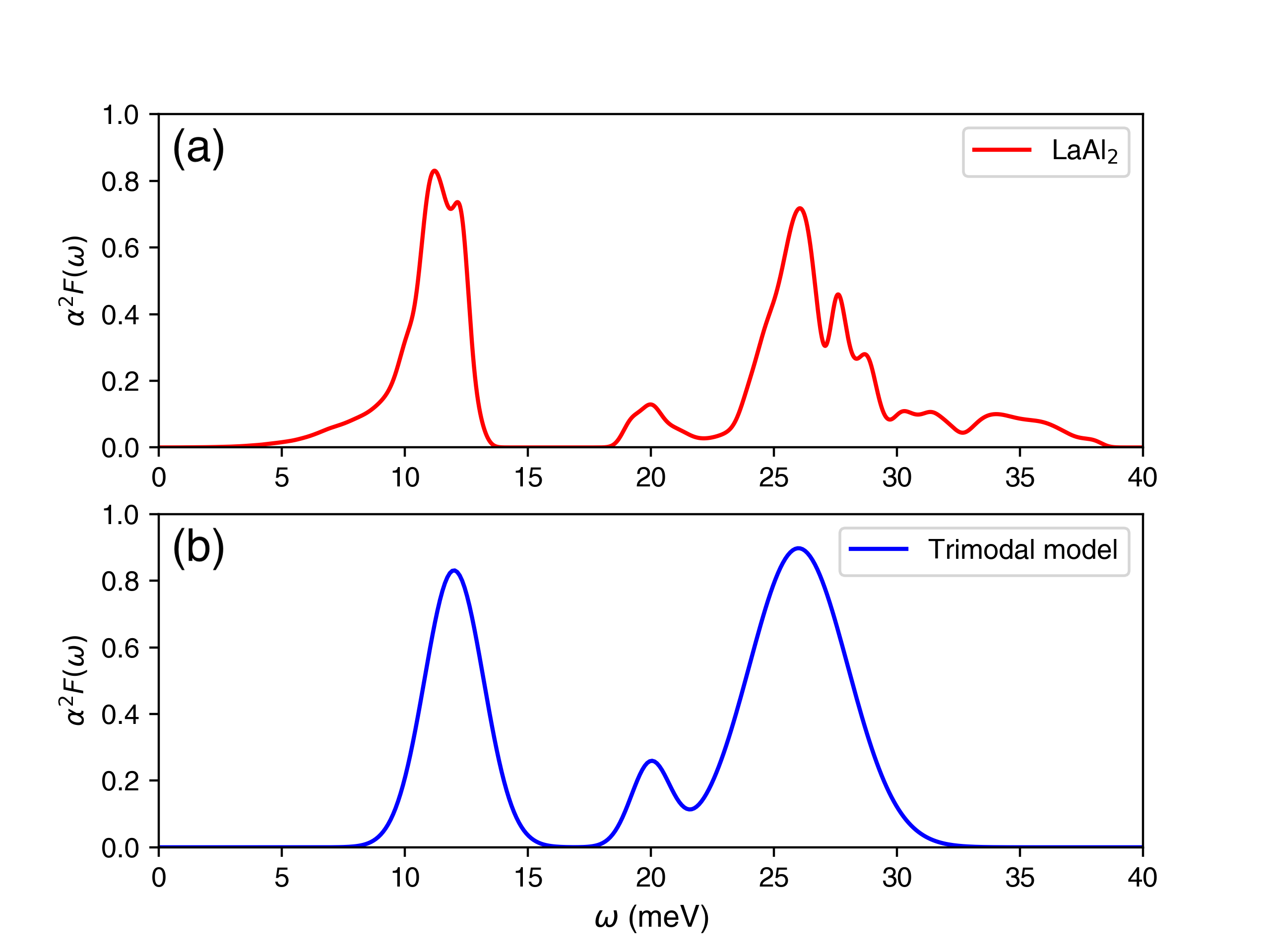}
  \caption{Comparison of (a) $\alpha^2F(\omega)$ for LaAl$_2$ with (b) a trimodal model $\alpha^2F(\omega)$ illustrates that the model spectral functions can resemble real materials.}
  \label{trimodala2f}
\end{figure}

The artificially generated $\alpha^2F(\omega)$ consist of three Gaussian peaks with randomly selected peak location and height,
\begin{eqnarray}
  \alpha^2F(\omega)=\sum_{i=1}^{  3}\frac{\lambda_i\omega}{2} g(\omega - \omega_i),
\end{eqnarray} 
where $g(\omega)$ is a normalized Gaussian with width of 1/8 of the peak frequency. 
%{\blue [RGH: We need to say something about the width of the Gaussians here. Also, are we really normalizing each Gaussian by $N$?]} %{\red The set that includes unimodal and bimodal distributions as well.}
The total $\lambda$ is then equal to the sum of the $\lambda_i$, which simplifies sampling of the space of spectral functions.
The artificial trimodal $\alpha^2F$s resemble the spectral functions of many realistic materials, see Fig.~\ref{trimodala2f} for the example of LaAl$_2$. The Allen-Dynes and Eliashberg $T_c$ for the hydrides are obtained from published work (see Refs.\ in Tab.~\ref{tab:data_summary}).

\begin{figure}[t]
  \includegraphics[width=\columnwidth]{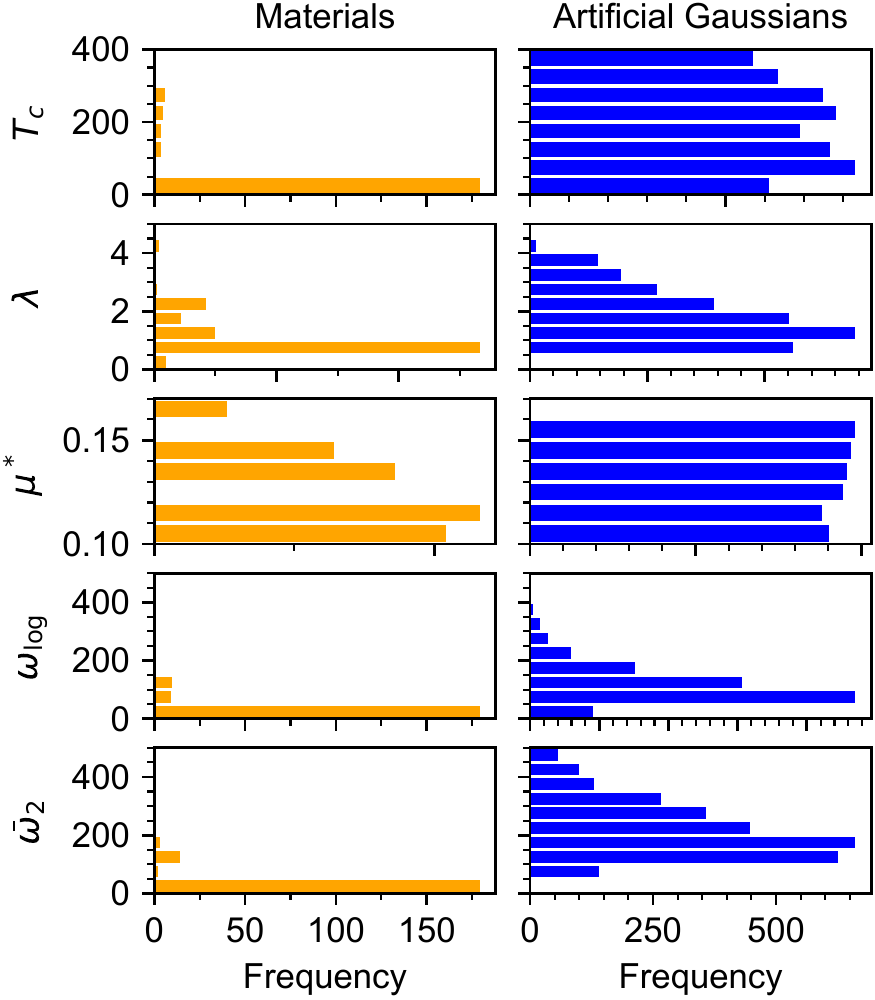}
  \caption{Histograms of input spaces of materials data (orange) and artificial Gaussian models (blue). Compared to the materials data, the artificial Gaussian models exhibit superior coverage of the input spaces. We generated artificial Gaussian models by sampling inputs uniformly with Sobol sequences and retaining entries with $T_c \leq 400$.}
  \label{fig:sobol_hist}
\end{figure}

To ensure efficient sampling of the input spaces, we select values of $\lambda$ and $\mu^\ast$ with pseudorandom Sobol sequences. As shown in Fig.~\ref{fig:sobol_hist}, our uniform sampling scheme results in a set of artificially generated $\alpha^2F$ corresponding to an approximately uniform distribution of $T_c$. Next, we removed artificial entries with $T_c > 400 K$ to better reflect the distribution of realistic materials. While the histogram of $\mu^\ast$ remains approximately uniform after this truncation, the histograms of $\lambda$, $\omega_{\log}$, and $\bar{\omega}_2$ become skewed towards lower values.

\subsection{Data}

\begin{table}[b]
    \caption{\label{tab:data_summary} Summary of the datasets used for training and validation of the machine learning model.}
    \begin{ruledtabular}
        \begin{tabular}{llcc}
        Name  & Entries\footnotemark[1] \footnotetext[1]{Unique/Resampled with varying $\mu^\ast$} & Training & Validation \\ \hline
        Calculated\footnotemark[2] \footnotetext[2]{This work} & 13/30 & Y  & \\
        Gaussian\footnotemark[2] & 2819/- & Y \\
        Literature\footnotemark[3] \footnotetext[3]{Published papers \cite{Uzunok2018,Singh2019,Arslan2016,Ttnc2015,Ttnc2016,Uzunok2016,Uzunok2017,Dilmi2018,Sichkar2013,Li2015,Ttnc2015b,Uzunok2017b,Uzunok2019,Karaca2016,Uzunok2018b,Ttnc2017,Ttnc2016b,Karaca2016b,Wang2017,Shrivastava2018b,Saib2017,Dabhi2016,Shrivastava2018,Wu2019,Karaca2019,Ttnc2014,Pawar2019,Bekaert2016,Acharya2017,Uzunok2020,Ttnc2012,Yue2018,AlcntaraOrtigoza2014,Cuamba2016,Chen2016,Ttnc2017b,Karaca2016c,Ttnc2019,Singh2018,Ttnc2018,Ono2020,doi:10.1021/acsami.8b17100,Krugloveaat9776}} & 29/149 & Y & \\
        Hydrides\footnotemark[3] & 13/19 & & Y \\
        \end{tabular}
    \end{ruledtabular}
\end{table}

\begin{figure}[t]
  \includegraphics[width=\columnwidth]{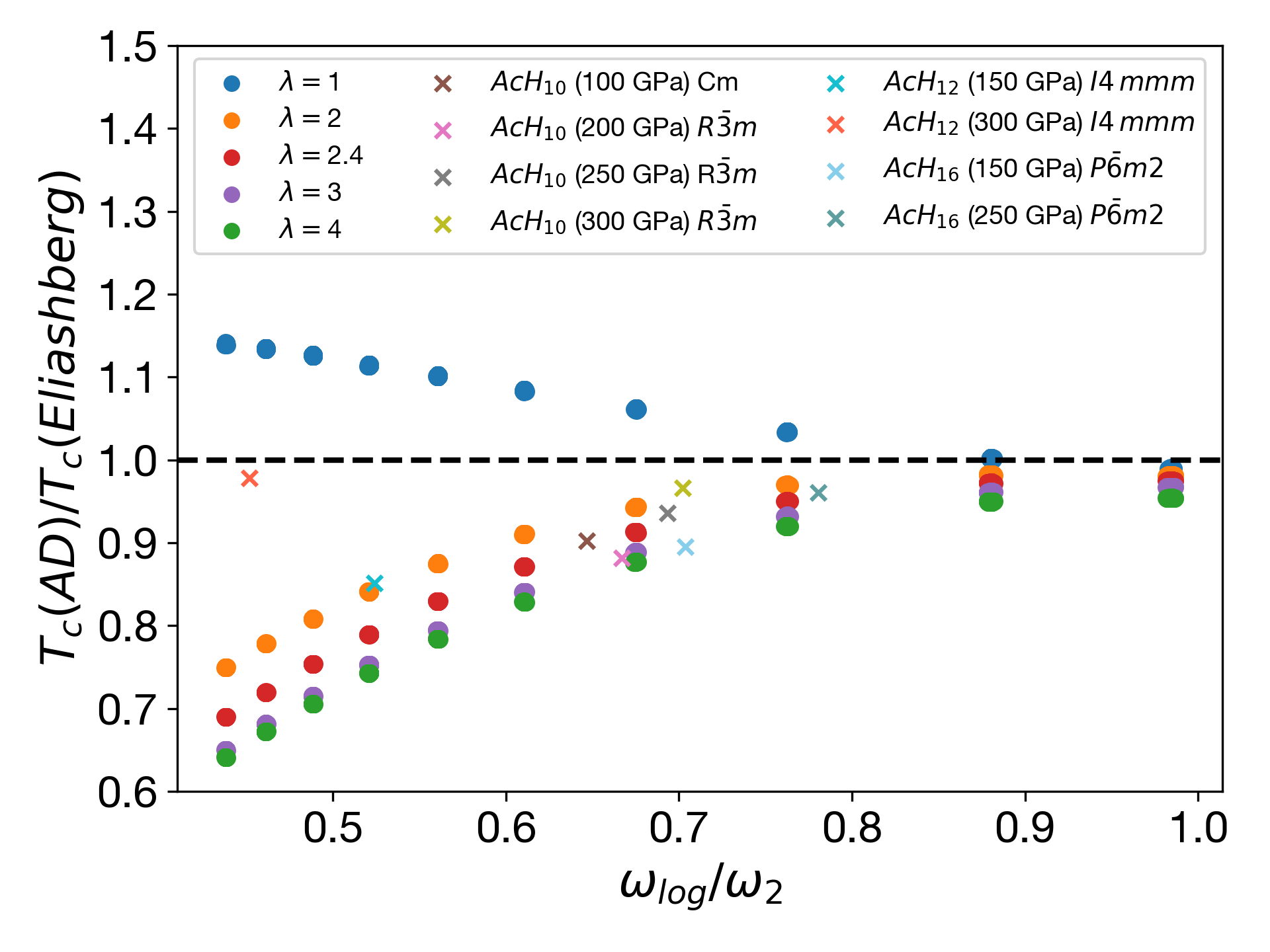}
  \caption{Ratio of the Allen-Dynes and Eliashberg $T_c$ for a bimodal Einstein-like model compared to data for hydrides are obtained from Refs.~\cite{doi:10.1021/acsami.8b17100,Krugloveaat9776}.  For the bimodal spectral functions, we select $\lambda_1 = \lambda_2$ for simplicity and vary the total $\lambda$ from 1 to 4.}
  \label{fig:ratio}
\end{figure}

In the Allen-Dynes formula, the “arbitrarily chosen” shape-dependent factor $f_2$  is based on the numerical solutions using the spectral functions of Hg, Pb, and the Einstein model~\cite{Allen-Dynes1975}. Because the number of $\alpha^2F(\omega)$ shapes is small, it is expected that the Allen Dynes $T_c$ ($T_c^{AD}$) would have significant errors in some instances. Fig.~\ref{fig:ratio} illustrates such deviations for bimodal Gaussian spectral functions.
%The corresponding frequency moments are $\expval{\omega_{\log}}=\omega_{1}^{\lambda_1/\lambda}\omega_{2}^{\lambda_2/\lambda}$ and $\expval{\bar{\omega}_2} = \sqrt{\frac{\lambda_1}{\lambda}\omega_{1}^2 + \frac{\lambda_2}{\lambda}\omega_{2}^2}$.
So far, we discussed the $\alpha^2F(\omega)$ shapes in an abstract sense because there is no single parameter that uniquely determines their shape. Allen and Dynes proposed using the ratio $\expval{\omega_{\log}}/\expval{\bar{\omega}_2}$ as an indicator of the shape of $\alpha^2F(\omega)$. In Fig.~\ref{fig:ratio}, the ratio $\frac{T_c^{AD}}{T_c^E}$ is plotted against $\expval{w_{\log}}/\expval{\bar{w}_2}$ for $\lambda$=0.6, 1, 2, 3 and 4. The results demonstrate that there can be significant differences between the Allen-Dynes $T_c^{AD}$ and Eliashberg $T_c^E$ even for some simple cases. The root mean square error ratio in the Allen Dynes paper is around 5.6$\%$ which we indicate by two horizontal dashed red lines in Fig.~\ref{fig:ratio}. When the ratio $\expval{w_{\log}}/\expval{\bar{w}_2}$ is 1, the shape of $\alpha^2F$ is that of the unimodal Einstein model and the Allen-Dynes $T_c$ accurately predict the Eliashberg $T_c$ regardless of the coupling strength. When the ratio $\expval{w_{\log}}/\expval{\bar{w}_2}$ decreases, i.e. the shape of $\alpha^2F$ has more structure; whether the Allen-Dynes formula can then still reasonably predict the Eliashberg $T_c$ depends on the electron-phonon coupling strength.

In this work, we train and test machine-learning models using the datasets listed in Tab.~\ref{tab:data_summary}. Two sizes are reported for each non-Gaussian dataset, indicating the number of unique materials compared to the total number of datapoints. We sample $\mu^\ast$ between [0.1, 0.16] which covers a wide range of possible $\mu^\ast$ values~\cite{ALLEN19831, Allen-Dynes1975}. The calculated, artificial Gaussian, and literature-derived $\alpha^2F$ datasets are used for training all machine learning models. We left the hydride materials out of the training in order to validate the extrapolative capacity of each model.

\subsection{Symbolic Regression}

As in our previous symbolic regression effort~\cite{Xie2019}, we use the SISSO framework to generate millions of candidate expressions by recursively combining the input variables with mathematical operators such as addition and exponentiation. Based on memory constraints, the subspace of expressions was limited to those generated within four iterations. This limitation precludes the appearance of expressions of the complexity of the Allen-Dynes equation, motivating our search for a dimensionless correction to the McMillan equation rather than directly learning models for $T_c$.

The initial quantities for generating expressions were the three dimensionless quantities $\lambda$, $\mu^\ast$, and the ratio $\omega_{\log}/\bar{\omega}_2$. Candidates were generated using the set of operators $\{+, -, \times, \/ \exp, \log, \sqrt{}, \sqrt[3]{}, \;^{-1}, \;^2, \;^3\}$. During the sure independence screening (SIS) step, these expressions were ranked based on their correlation to the ratio $T_c^{\text{E}}/T_c^{\text{McMillan}}$ rather than $T_c^{\text{E}}$ to identify dimensionless, multiplicative corrections to $T_c^{\text{McMillan}}$.

To facilitate generalizability, we employ leave-cluster-out cross-validation during the generation of expressions using $k$-means-clustering with $k = 10$ on the combined set of 179 non-hydride and 2819 artificial-Gaussian entries. For each round of cross-validation, we generate candidate equations using a different subset of nine clusters and used the remaining cluster to evaluate performance using the root-mean-square error metric. As such, each training sample was left out of training and used for testing during one round. The top 10,000 models, ranked by root-mean-square error (RMSE) across the training set, were returned from each round. Models that did not appear in all ten rounds, corresponding to those with poor performance in one or more clusters, were eliminated. Following the same principle, we ranked the remaining equations by the average RMSE across all ten rounds. A selection of candidate equations and their RMSE is available in the supplemental information.

We note that the Sparsifying Operator (SO) step of the SISSO framework offers increased model complexity, as we explored in our previous work, but is limited in functional form to linear combinations of expressions generated from the preceding step. The linear combination of expressions from the initial subspace, by extension, also excludes equations as complex as the Allen-Dynes correction. Therefore, we did not consider linear combinations of expressions, meaning the SO simply selected the first-ranked expression from the SIS step in each run.

%{\blue [RGH: This paragraph seems to miss a conclusion, such that we therefore do not perform the SO step.]}

\section{Results}
\subsection{Correction factors for $T_c$ from symbolic regression}

\begin{figure}[t]
  \includegraphics{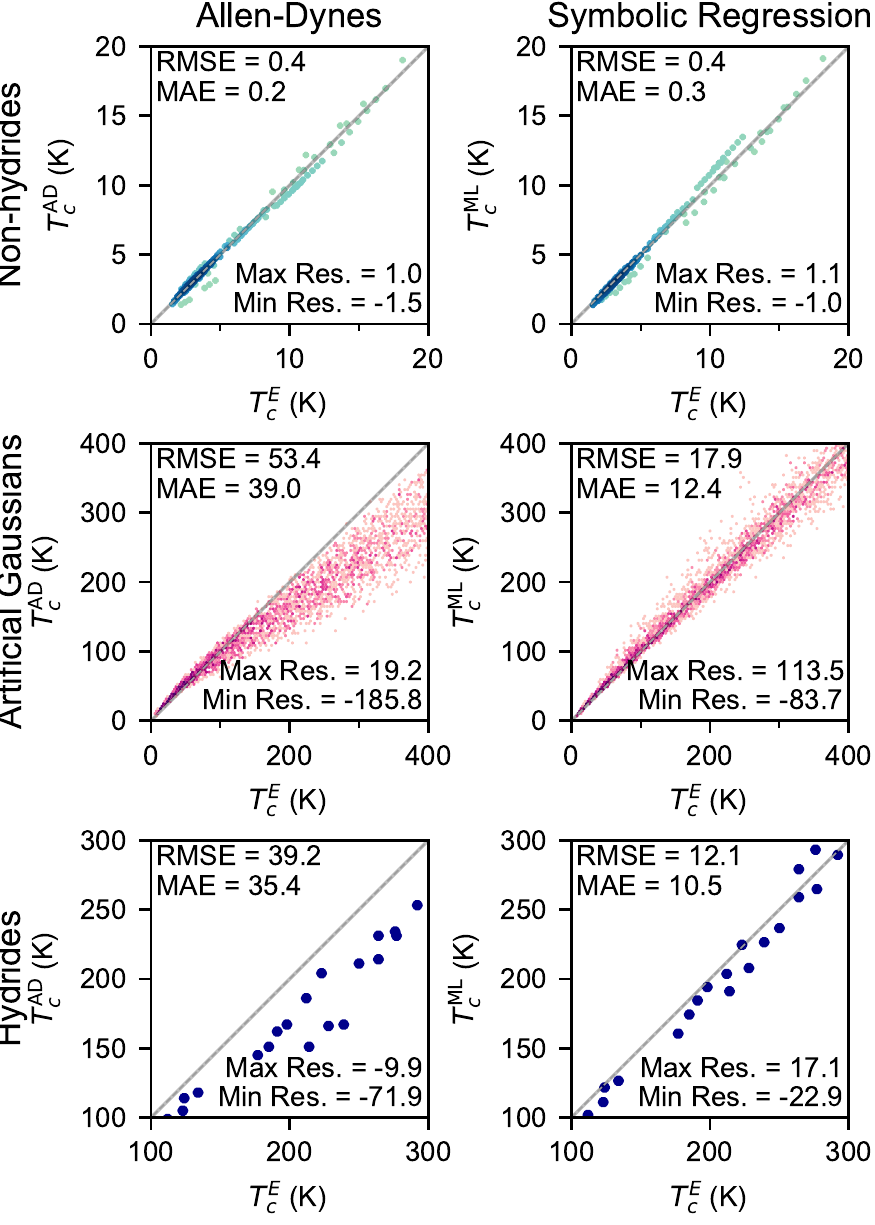}
  \caption{Comparison of predictions using the Allen-Dynes equation (left column) and the new symbolic-regression corrections (right column). $T_c^{\text{E}}$ is plotted against $T_c^{\text{Model}}$ such that accurate predictions lie on the gray 1:1 line. Non-hydride (top row) and artificial Gaussian (middle row) panels depict the training error while the hydride~\cite{doi:10.1021/acsami.8b17100,Krugloveaat9776}(bottom row) panels show extrapolative capacity. The non-hydride and artificial Gaussian panels are colored by the log-density of points. We report the root-mean-square error (RMSE), mean-absolute error (MAE), maximum residual, and minimum residual values in Kelvin. The maximum residual corresponds to the largest overprediction while the minimum residual corresponds to the largest underprediction. The two multiplicative factors obtained from symbolic regression improves the prediction compared with the two multiplicative factors of the Allen-Dynes formula, particularly for the higher $T_c$ systems.}
  \label{fig:rmse}
\end{figure}

We performed symbolic regression twice, sequentially, to obtain two dimensionless prefactors of the McMillan exponential, yielding a machine learned critical temperature,
\begin{equation}
T_c^{\textrm{ML}}= \frac{f_\omega f_\mu\, \omega_{\log}}{1.20}\exp\left(-\frac{1.04(1+\lambda)}{\lambda-\mu^{\ast}(1+0.62\lambda)}\right).
\label{eq:MLTc}
\end{equation}
We name the two learned prefactors {\it a posteriori} based on their functional forms and the mechanisms by which they reduce the error in predicting $T_c$. The first factor,
\begin{equation}
f_\omega = 1.92\left(\frac{\lambda + \frac{\omega_{\log}}{\bar{\omega}_{2}} -\sqrt[3]{\mu^{\ast}}}{\sqrt{\lambda}\exp(\frac{\omega_{\log}}{\bar{\omega}_{2}})}\right) - 0.08
\label{eq:f_omega}
\end{equation}
is obtained from the fit to the ratio $T_c^{\text{E}}/T_c^{\text{McMillan}}$ and eliminates the systematic underprediction of $T_c$ at higher temperatures. Like the Allen-Dynes prefactor $f_2$, $f_\omega$ includes the ratio $\omega_{\log}/\omega_2$, modifying the prediction based on the shape of $\alpha^2F(\omega)$. Moreover, $f_\omega$ also scales with $\sqrt{\lambda}$, like the Allen-Dynes prefactor $f_1$.  This is in agreement with the correct large-$\lambda$ behavior of Eliashberg theory, unlike our earlier work~\cite{Xie2019} and the modified $T_c$ equation with linear correction proposed recently by Shipley et al.~\cite{Shipley2021}.  The manifestation of both behaviors in $f_\omega$ gives credence to our symbolic regression approach because it incorporates the primary effects of the Allen-Dynes equation with fewer parameters. Applying the correction $T_c = f_\omega T_c^{\text{McMillan}}$ achieves a percent RMSE of 15.2\% across the materials (non-Gaussian model) data, compared to 48.6\% when using the Allen-Dynes equation.

%\begin{equation}
%f_\omega = 1.7156\left(\frac{\lambda + \frac{\bar{\omega}_{2}}{\omega_{\log}} -\sqrt{\mu^{\ast}}}{\sqrt{\lambda}\exp(\frac{\bar{\omega}_{2}}{\omega_{\log}})}\right)
%\end{equation}

The second correction factor
\begin{equation}
f_\mu = \frac{6.86\exp\left(\frac{-\lambda}{\mu^\ast}\right)}{\frac{1}{\lambda} - \mu^\ast - \frac{\omega_{\log}}{\bar{\omega}_{2}}} + 1
\label{eq:f_mu}
\end{equation}
is obtained from the fit to the ratio $T_c^{E}/(f_\omega T_c^{\text{McMillan}})$, effectively correcting the residual error from the fit of $f_\omega$ and thus cannot be used independently. Applying the correction $T_c = f_\omega f_\mu T_c^{\text{McMillan}}$ achieves a percent RMSE of 15.1\% across the materials datasets, compared to 15.2\% when using $f_\omega$ alone. The influence of $f_\mu$ is more apparent when examining clusters of points corresponding to resampled $\mu^\ast$ values for a single material, where the systematic error in $T_c^{\text{ML}}/T_c^{\text{E}}$ is reduced.
%\begin{equation}
%f_\mu = 0.5317\left(e^{-\mu^{\ast}} - e^{-\lambda} - \frac{\mu^{\ast}+\lambda}{\mu^{\ast}-\lambda}\right)
%\end{equation}
 Note that $f_\mu\rightarrow 1$ in both of the limits $\lambda\rightarrow 0$ and $\lambda\rightarrow\infty$, and in fact does not vary by more than $\sim 10\%$ from 1 over the data set.

Fig.~\ref{fig:rmse} shows that, apart from the low-$T_c$ non-hydride materials for which the difference is smaller than 0.1~K, the corrections $f_\omega$ and $f_\mu$ dramatically improve predictions compared to using the Allen-Dynes equation. Since we excluded the hydrides from the training, these results successfully validate our data-driven symbolic regression approach by demonstrating the extrapolative capacity of the learned equations.

To further quantify the similarity between the existing Allen-Dynes prefactors and the machine-learned prefactors, we employ two statistical measures, the Spearman and distance correlation. The Spearman correlation is a measure of monotonicity in the relationship between rankings of two variables. Like the Pearson correlation coefficient for linear correlation, the Spearman correlation varies between $-1$ and +1, where extrema imply high correlation and zero implies no correlation. Unlike the Pearson correlation, the Spearman correlation does not assume normally distributed datasets. By construction, all four prefactors tend to one for many materials, resulting in asymmetric distributions that are unsuitable for analysis with parametric measures like the Pearson correlation.

In addition to the Spearman correlation, we compute the distance correlation, another nonparametric measure of the dependence between two variables. The distance correlation is defined as the ratio of the distance covariance and the product of the distance standard deviations, where distance covariance is the weighted Euclidean distance between the joint characteristic function of the two variables and the product of their marginal characteristic functions. Unlike the Pearson and Spearman correlation coefficients, the distance correlation varies between 0 and 1, where 0 indicates that the variables are independent, measuring both linear and nonlinear association. 

Tab.~\ref{tab:correlation} shows a strong relationships between $f_1$, $f_2$, and $f_\omega$ according to both Spearman and distance correlation metrics, with values close to one. This numerical analysis reinforces the conclusion that $f_\omega$ reproduces characteristics of both $f_1$ and $f_2$, as illustrated earlier in the comparison of functional forms. On the other hand, both Spearman correlation and distance correlation measures indicate slightly weaker relationships between $f_\mu$ and the other three prefactors. The relative independence of $f_\mu$ compared to $f_\omega$, $f_1$, and $f_2$ stems from the sequential nature of the fitting process.

\begin{table}[t]
    \caption{\label{tab:correlation} Correlations between Allen-Dynes and machine-learned prefactors. The Spearman correlation measures the rank correlation between two variables while the distance correlation measures the dependence between two variables, including both linear and nonlinear association. Values closer to zero indicate weaker relationships. While $f_1$, $f_2$, and the new prefactor $f_\omega$ are strongly correlated to one another, $f_\mu$ shows a weaker relationship. The relative independence of $f_\mu$ reflects its origin as a second correction fit to the residual error of $f_\omega$.}
    \begin{ruledtabular}
        \begin{tabular}{llrr}
        & & \multicolumn{2}{c}{Correlation}\\ \cline{3-4}
            \multicolumn{2}{c}{Prefactors} & Spearman & Distance \\ \hline
            $f_1$      & $f_2$      &  0.932    & 0.964    \\
            $f_1$      & $f_\omega$ &  0.943    & 0.973    \\
            $f_1$      & $f_\mu$    & -0.870    & 0.693    \\
            $f_2$      & $f_\omega$ &  0.850    & 0.965    \\
            $f_2$      & $f_\mu$    & -0.861    & 0.662    \\
            $f_\omega$ & $f_\mu$    & -0.887    & 0.731   
        \end{tabular}
    \end{ruledtabular}
\end{table}

\subsection{Comparing predictive models for $T_c$}
\begin{figure}[t]
  \includegraphics{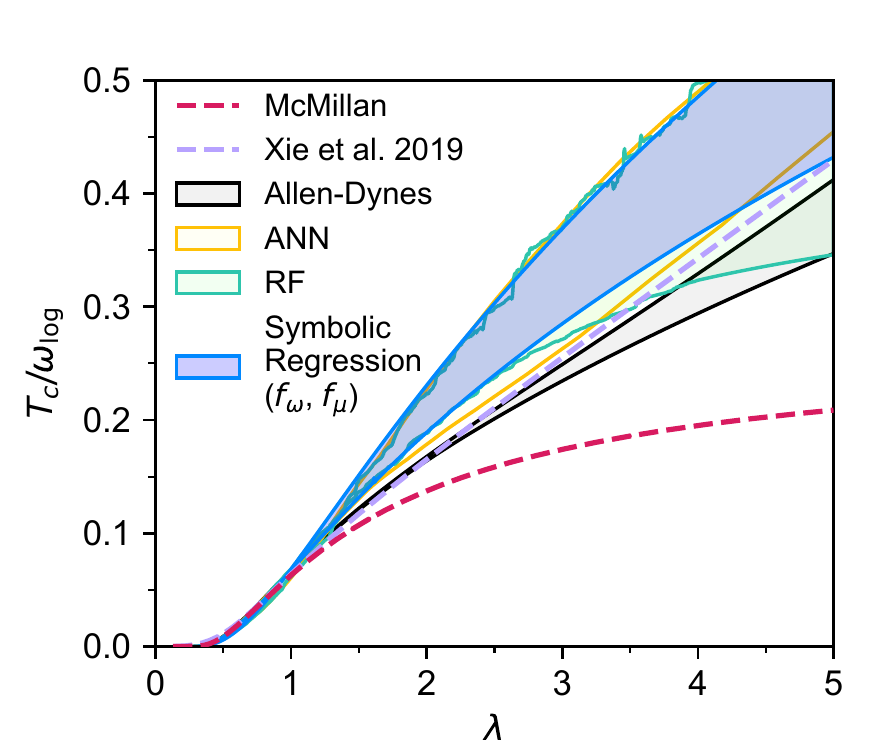}
  \caption{Dependence of $T_c$ on $\lambda$ for select predictive models. The McMillan and Xie at al. (2019) equations, which do not change with $\omega_{\log}/\bar{\omega}_2$, are depicted as dashed curves. The Allen-Dynes formula and the ANN, RF, and symbolic regression machine learning corrections from this work are plotted as shaded regions bound by $\omega_{\log}/\bar{\omega}_2=1.1$  and $\omega_{\log}/\bar{\omega}_2=1.6$ curves. All models behave similarly for low to moderate values of $\lambda$. For larger values of $\lambda$, the ANN, RF, and symbolic regression corrections deviate significantly from the Allen-Dynes equation as well as the previous symbolic regression equation.}
  \label{fig:lambda}
\end{figure}

To compare existing equations for $T_c$ with the corrections identified in this work, we benchmarked the RMSE across non-hydride materials, artificial Gaussians, and hydrides as tabulated in Table \ref{tab:benchmark}. Additionally, we compute the \%RMSE by normalizing each RMSE by the mean value across the corresponding dataset. To assess the behavior of each model with increasing $\lambda$, we plot $T_c/\omega_{\log}$ for each model in Fig. \ref{fig:lambda}.

As expected, the Allen-Dynes equation improves on the McMillan equation across all three groups. On the other hand, the equation identified by Xie et al.\cite{Xie2019} in an earlier symbolic regression work performs slightly worse on the low-$T_c$ non-hydride dataset but achieves lower RMSE across the artificial Gaussian and hydride materials despite being trained on a small set of 29 low-$T_c$ materials. 

\begin{table*}[tb]
    \caption{\label{tab:benchmark} Comparison of model performance on materials and artificial Gaussian datasets.}
    \begin{ruledtabular}
        \begin{tabular}{lrrrrrr}
        & \multicolumn{6}{c}{Error relative to $T_c^{\text{E}}$}\\ \cline{2-7}
        & \multicolumn{2}{c}{Non-hydride materials} & \multicolumn{2}{c}{Artificial Gaussians} & \multicolumn{2}{c}{Hydrides} \\ % \cline{2-3} \cline{4-5} 
            Model & RMSE (K) & \%RMSE & RMSE (K) & \%RMSE & RMSE (K) & \%RMSE\\ \hline
            McMillan~\cite{Mcmillan1968} & 0.8 & 14.4 & 88.1 & 45.1 & 76.8 & 36.6\\
            Allen-Dynes~\cite{Allen-Dynes1975} & 0.4 & 6.5 & 53.4 & 27.3 & 39.2 & 18.7\\
            Xie et al. 2019~\cite{Xie2019} & 0.9 & 16.7 & 36.7 & 18.8 & 25.7 & 12.3\\ 
            Symbolic Regression ($f_\omega$)\footnotemark[1] \footnotetext[1]{This work} & 0.5 & 8.4 & 17.9 & 9.2 & 12.2 & 5.8\\
            Symbolic Regression ($f_\omega$, $f_\mu$)\footnotemark[1] & 0.4 & 7.3 & 17.9 & 9.2 & 12.1 & 5.8\\
            Random Forest (RF)\footnotemark[1]& 0.2 & 3.3 & 13.7 & 7.0 & 9.9 & 4.7\\
            Artificial Neural Network (ANN)\footnotemark[1] & 0.2 & 4.0 & 17.0 & 8.7 & 17.4 & 8.3 \\
        \end{tabular}
    \end{ruledtabular}
\end{table*}

Applying the new $f_\omega$ prefactor to the McMillan equation reduces \%RMSE in non-hydride materials from 14.4\% to 8.4\%, in artificial Gaussian models from 45.1\% to 9.2\%, and in hydrides from 36.6\% to 5.8\%. Moreover, applying both $f_\omega$ and $f_\mu$ results in a further, modest improvement to the RMSE. In Fig.~\ref{fig:lambda}, our machine-learned correction (blue) is nearly equal to the Allen-Dynes equation (gray) for values of $\lambda$ up to 1 but rapidly increases at larger $\lambda$. Both bounds, for higher and lower values of $\omega_{\log}/\bar{\omega}_2$, exceed the bounded region of the Allen-Dynes equation, indicating that at least part of the new model's success is due to an improvement in capturing the behavior of $T_c$ with increasing $\lambda$. 

We additionally fit a random forest (RF) model and an artificial neural network (ANN) model using the same training data to compare against our symbolic regression method. Hyperparameters for RF and ANN models were selected using 10-fold leave-cluster-out cross-validation and the same clusters identified for symbolic regression. On the other hand, the model error was estimated using nested cross-validation, where the inner loop was performed using a conventional 5-fold cross-validation scheme. Production models used in Fig.~\ref{fig:lambda} were fit with the selected hyperparameters, available in the supplemental information, using the entire training set.

The RF is an ensemble model comprised of decision trees, each fit to random subsets of the data and queried to yield an independent prediction. Each decision tree uses a flow-chart-like series of decisions (branches) to yield predictions (leaves) and is optimized by varying decision thresholds. While individual decision trees are prone to overfitting, a random forest produces robust predictions by averaging the predictions of its members. The optimized RF model, consisting of 100 decision trees with a maximum depth of eight splits per tree, achieved the lowest RMSE across all three models, with 4.7\% RMSE in the testing set of hydride materials. This success may be attributed to both the flexibility of the method and the relative complexity compared to other methods. With up to 128 nodes per tree, the RF evaluates tens of thousands of binary decisions per prediction. On the other hand, as illustrated in Fig.~\ref{fig:lambda}, the resulting output (green) is discontinuous. Furthermore, the RF does not have the ability to extrapolate outside of regions of the input spaces included in the training data, resulting in constant-value outputs. This deficiency is evident in both upper- and lower-bound curves above $\lambda = 3.8$, where the RF correction results in a simple rescaling of the McMillan curve.

The ANN models in this work are feedforward neural networks, also known as multi-layer perceptrons, designed to learn highly non-linear function approximators to map multiple inputs to a target output. The feedforward architecture involves an input layer consisting of one neuron per input, one or more hidden layers, and an output layer consisting of one neuron per target. The value at each non-input neuron is a weighted, linear summation of the values in the preceding layer followed by a non-linear activation function. The optimized ANN includes three hidden layers with forty neurons each, totaling 3,521 trainable parameters of multiplicative weights and additive biases. Despite the increased model complexity, the ANN performs similarly to the symbolic regression model, with slightly lower training RMSE and slightly higher testing RMSE. With increasing $\lambda$, the ANN model yields similar values of $T_c$ as indicated by the overlap between the shaded regions of the symbolic regression model (blue) and the ANN (yellow).

For low to moderate values of $\lambda$, such as those originally studied by Allen and Dynes, all models behave similarly and the dimensionless corrections ($f_1$, $f_2$, $f_\omega$, $f_\mu$, ANN, RF) are close to unity. However, as $\lambda$ increases, the ANN, RF, and symbolic regression corrections deviate significantly from the Allen-Dynes equation as well as the previous symbolic regression equation~\cite{Xie2019}. The corrections introduced in this work successfully correct the systematic underprediction of $T_c$, with the symbolic regression solution offering simplicity and accuracy. Moreover, the monotonicity constraint in the symbolic regression search guarantees invertibility, allowing experimentalists to extract $\lambda$ from measured $T_c$ and $\omega_D$. This characteristic is not guaranteed for the RF and ANN models.

\section{Summary}

The present work demonstrates the application of symbolic regression to a curated dataset of $\alpha^2F(\omega)$ spectral functions, yielding an improved analytical correction to the McMillan equation for the critical temperature of a superconductor. We showed that the well-known Allen-Dynes equation, 
an  early improvement based on fitting to a very limited set of spectral functions, exhibits systematic error when predicting the Eliashberg  critical temperature of high-$T_c$ hydrides, a flaw due to the original training set based on low-$T_c$
superconductors.  The equation we obtain here by symbolic regression has the same form as the original Allen-Dynes equation, with exactly the same MacMillan exponential factor, but has two prefactors that behave
very differently than those employed by Allen-Dynes.  They ensure that  superconductors with spectral functions, $\alpha^2F(\omega)$, of
unusual shapes, such that $\bar\omega_2/\omega_{\log}$ is significantly different from 1, are adequately
described; this subset of conventional superconductors
includes the new hydride high-pressure superconductors.
In addition, the machine-learned equation can be simplified by dropping one of the prefactors with negligible loss of accuracy. 
% We propose that this equation should replace the Allen-Dynes equation for prediction of higher-temperature superconductors and 
% for the estimation of $\lambda$ from experimental data.
Since the machine-learned expression of Eqs.~\eqref{eq:MLTc}-\eqref{eq:f_mu} extends the accuracy of the Allen and Dynes expression to high-temperature superconductors while maintaining the utility and simplicity of the original formula, we suggest that this equation should replace the Allen-Dynes formula for predictions of critical temperatures and estimations of lambda from experimental data, particularly for higher-temperature superconductors.

Using a dataset of \emph{ab initio} calculations alongside artificially-generated spectral functions, we mitigated 
the small-data problem associated with previous symbolic-regression efforts. The dimensionless correction factor identified by symbolic regression reproduces the expected physical behavior with increasing $\lambda$ and achieves lower prediction errors than the Allen-Dynes corrections, despite having similar model complexity. Finally, we compared our equation to models generated with two other machine-learning techniques, which achieve modest improvements in error at the cost of far greater complexity and lack of invertibility. While the present work successfully learns the isotropic Eliashberg $T_c$, future extensions may incorporate additional data from fully-anisotropic Eliashberg calculations and experimental measurements. On the other hand, separate extensions may involve approximating $\alpha^2F$-related quantities from less-expensive calculation of density functional theory-based descriptors like the electronic density of states.

\begin{acknowledgements}
We are grateful to L. Boeri, P. Allen, and W. Pickett for valuable discussions. We thank Dmitrii Semenok for providing data related to Actinium hydrides. The work presented here was performed under the auspice of Basic Energy Sciences, United States Department of Energy, contract number DE-SC0020385. Partial funding was also provided by the University of Florida Informatics Institute.
\end{acknowledgements}

\section{Data Availability}

The database of computed spectral functions $\alpha^2F(\omega)$, the derived descriptors, and the critical temperatures are freely available at \url{https://MaterialsWeb.org} and Github (\url{https://github.com/henniggroup/}). The symbolic regression workflow software we developed is freely available on Github as well.

\section{Author Contributions}
All authors contributed extensively to the work presented in this paper. SRX, YQ, ACH, LF, JL, JK, GRS, JJH, PJH, and RGH conceived the overall methodology of data assembly, augmentation, Eliashberg calculations, and symbolic regression. YQ, BD, JMD, IS, and USH performed the literature search and collected the spectral function data for superconductors. SRX and YQ implemented the algorithm and performed the calculations and analysis. SRX, YQ, PJH, and RGH contributed to the writing of the manuscript.

\section{Competing Interests statement}
The Authors declare no Competing Financial or Non- Financial Interests.

\bibliographystyle{apsrev4-1} 
\bibliography{references}

\end{document}